\documentclass[a4paper,12pt]{article}
\usepackage{epsfig}
\usepackage{amssymb}
\begin{document}
\thispagestyle{empty}

\newcommand{\etal}  {{\it{et al.}}}  
\def\Journal#1#2#3#4{{#1} {\bf #2}, #3 (#4)}
\def\PRD{Phys.\ Rev.\ D}
\def\NIMA{Nucl.\ Instrum.\ Methods A}
\def\PRL{Phys.\ Rev.\ Lett.\ }
\def\PLB{Phys.\ Lett.\ B}
\def\EPJ{Eur.\ Phys.\ J}
\def\IEEETNS{IEEE Trans.\ Nucl.\ Sci.\ }
\def\CPCD{Comput.\ Phys.\ Commun.\ }



{\Large\bf
\begin{center}
To what extent should we find the chiral end point?
\end{center}
}

\begin{center}
\large{ G.A. Kozlov  }
\end{center}
\begin{center}
 { Bogolyubov Laboratory of Theoretical Physics\\
 Joint Institute for Nuclear Research,\\
 Joliot Curie st., 6, Dubna, Moscow region, 141980 Russia }
\end{center}


 \begin{abstract}
 \noindent
 {We analyze the critical phenomena in the  theory of strong interactions at high temperatures starting from first principles.  The model is based on the dual Yang-Mills theory with scalar degrees of freedom - the dilatons. The latter are produced due to the spontaneous breaking of an approximate scale symmetry. The phase transitions are considered in systems where the field conjugate to the order parameter has the (critical) chiral end mode. The chiral end point (ChEP) is a distinct singular feature existence of which is dictated by the chiral dynamics. The physical approach the effective ChEP is studied via the influence fluctuations of two-body Bose-Einstein correlation function for observed particles to which the chiral end mode couples.}


\end {abstract}




\bigskip

{\bf Introduction.-}  The new heavy-ion collisions facilities named as NICA and FAIR are awaiting to be completed in 2017-2018 in JINR (Dubna) and GSI (Darmstadt), respectively. Among the main aims of both the collider NICA and the accelerator with a fixed target FAIR there are the search of strong interacting matter at high temperatures, $T$, and baryon densities corresponding to the freezeout point for a given experiment. The freezeout point in particle physics is often referred to as critical point of quantum chromodynamics (QCD) or to as chiral end point (ChEP), which are different from the nuclear critical point.  At high  $T$ and (in general) at finite baryon densities, where non-vanishing baryon chemical potentials $\mu_{B}$ are assumed, the  matter becomes weakly coupled and at the vicinity of ChEP the color is no more confined, the chiral symmetry is restored. The phase transitions in the proximity of the ChEP are associated with breaking of symmetry. The ChEP itself may be clarified through the search of its location on the $(\mu_{B}-T$) plane, i.e. through the exploration of  $(\mu_{B}-T$) phase diagram (see, e.g., [1] and the refs. therein).

A few questions arise: what does it mean - the ChEP?  What the main observables would be measured to indicate the ChEP is really achieved? What are new knowledges one can accumulate if ChEP will be approached? To realize the program of study the critical phenomena (e.g., to predict the ChEP location on the phase diagram above mentioned) it is often advocated  that the collision energy in relativistic heavy ion machines would not be sufficiently high ($\sqrt {{s}_{NN}}\simeq$ 4 -11 GeV), however the new phenomena would be seeing at high luminosity  $\sim 10^{27}$ $cm^{-2}s^{-1}$, and the finite baryon density. 

One of the milestones for QCD to describe the strong interactions successfully, in contrast to classical non-Abelian gauge theory, is the manifestation of chiral symmetry breaking.
At large distances, QCD exhibits nonperturbative phenomena such as chiral symmetry breaking and confinement of color charges. The relation between these phenomena is not yet clarified in the frame of QCD, thus the correlation between phase transitions of chiral symmetry restoration and deconfinement in QCD at finite temperatures is an important issue.

The analytical treatment of QCD at high $T$ is essential for the understanding of the phenomena above mentioned, however it is very difficult in the sence of its realization due to massless quarks in the theory.  One of the intrinsic approaches to analytical calculations in strong interacting matter pattern is through the scheme with topological defects which emerge in some effective models.
The general properties of field models containing the defects are related to the fact that these defects exist only in the phase with spontaneously broken symmetry, where the average expectations of scalar fields emerge. In the phase with nonbroken symmetry the solutions describing the topological defects are absent.

The minimal model where the topological defects (strings)  arise is the Abelian Higgs-like model [2]. The key point is reducing of $SU(N)$ gluodynamics to $[U(1)]^{N-1}$ dual Abelian scalar theory. The breaking of the latter group gauge symmetry is realized through the Higgs-like mechanism.
The maximally Abelian gauge suggests the special properties of QCD vacuum, such as Abelian dominance and the condensation of scalars [3] which provide the dual superconductor picture of the QCD vacuum [4]. In this vacuum the color-electric flux is squeezed into an almost one-dimensional object such as string due to the dual Meissner effect caused by scalar condensation. This is the dual analogy of the Abrikosov vortex in the classical superconductor supported by the Cooper pair condensation.

The close relation between chiral symmetry breaking and confinement has been suggested in terms of color-magnetic monopoles [5,6] which topologically appear in QCD by taking the maximally Abelian gauge. In the latter case, QCD becomes Abelian-like due to off-diagonal gluon mass of about  1 GeV. In lattice QCD by removing the monopole degrees of freedom (d.o.f.) the chiral symmetry breaking and confinement are lost simultaneously. This features the special role of a monopole to both chiral symmetry breaking and confinement which lead to the fact that two  QCD nonperturbative phenomena are related to each other through the monopoles. The essential point here is the description of long-distance Yang-Mills (YM) theory by a dual gauge theory in which all particles become massive via a dual monopole mechanism.

In this paper we  study the critical phenomena in strong interacting matter on a purely qualitative  level of understanding the non-perturbative gauge theory dynamics from first principles. We deal in the framework of an effective Abelian model of $SU(3)$ gluodynamics which allows to describe the infra-red (IR) properties of the vacuum, and where an essential point is an appearance of scalar d.o.f, the dilatons, in the theory. The dilaton - the pseudo Goldstone boson - is associated with spontaneous breaking of scale (or other global) symmetry of some four dimensional gauge theory. 

The approach to the QCD critical phenomena is supposed to be done  via the influence fluctuations of two-body Bose-Einstein (BE) correlation function for observed particles to which the chiral end mode couples. By studying BE correlations of identical particles (e.g., like-sign  charged particles of the same sort) or even off-correlations with respect to different charged bosons, one can predict and even experimentally determine the time and the spatial region (decoupling surface) over which particles do not have the interactions. However, for an evolving system due to ions (particles) collisions, there is no a surface, since at each time there is a spread out surface due to (thermal) fluctuations in the final interactions and correlations. Thus, the shape of the surface evolves even in time and the particle emitted source is not approximately constant. In paper [7] we have already carried out the extended model of BE correlations at finite temperature approach to quantum field theory which can be applied to experimental data. 
One of the parameters of the model is the effective temperature of the particle emission source under a random  (operator) force  influence. 

We propose a model in which the approximate scale invariance is manifest at high energies, but is spontaneously broken at a scale $f$ close to the QCD scale. The dilaton  is lighter than the other resonances which can have the masses $\sim 4\pi f$. We emagine that the entire structure of QCD is embedded in the conformal sector at high energies. Based on the dual YM theory, we study the effect of ChEP on two-particle BE correlation function $C_{2}$  of fluctuations of the correlation strength which can be extracted from experimental data. The quark-antiquark bound states are presented in terms of flux tubes the properties and correlations of which are under consideration. We suppose a strongly interacting medium is created in heavy ion  collisions which can exhibit the space momentum correlations and collective behavior. One of the essential quantities that influences the particle freezeout is the formation time in the flux tube fragmentation. Note that one or more of characteristics of $C_{2}$ function must be singular at the transition point, the ChEP.
The clear signature is non-monotonous behavior of $C_{2}$ which is sensitive to the proximity of ChEP and could be measured by the magnitude of the fluctuation length. 


{\it\bf Gauge field correlator.-} One of the features of ChEP is a fluctuation measure related to some observables which may be visible through the fluctuations of some characteristic length as a derivative of the chiral end mode. We explore the model describing the fluctuations based on the probability distribution of an order parameter field. To realize the program with estimation of the length and its fluctuations one has to deal with gauge-invariant quantities. The gauge-invariant two-point correlation function (TPCF) describing the correlator of two gluon field strengths $F_{\mu\nu}$ in QCD at different (Euclidean) space-time points, connected by a Schwinger color string 
$$ U_{A}(x,y) = P\,\exp \left [i\,e\,\int_{y}^{x}\,dz^{\mu}\,A_{\mu} (z)\right ]$$
has the form [8]
\begin{equation}
\label{e1}
T_{\mu\nu\lambda\rho} (x) =\langle e{^2}\,F_{\mu\nu} (x)\,U_{A}(x,0)\,F_{\lambda\rho} (0)\,U_{A} (0,x)\rangle 
\end{equation}
with the following Lorentz decomposition
$$\left (\delta_{\mu\lambda}\delta_{\nu\rho} - \delta_{\mu\rho}\delta_{\nu\lambda}\right )e^{2} \,D_{1}(x^{2}) + \frac{1}{2}\left [\partial_{\mu}\left (x_{\lambda}\delta_{\nu\rho} - x_{\rho}\delta_{\nu\lambda}\right ) + 
\partial_{\nu}\left (x_{\rho}\delta_{\mu\lambda} - x_{\lambda}\delta_{\mu\rho}\right )\right ]e^{2}\,D_{2} (x^{2}), $$
where $F_{\mu\nu} = \partial_{\mu} A_{\nu} - \partial_{\nu} A_{\mu} - i\,e [A_{\mu},A_{\nu}]$,  
$A_{\mu} (x)= \sum_{a} A^{a}_{\mu}(x)\,t_{a}$, $A^{a}_{\mu}(x)$ are the YM fields, $t_{a}$ are the generators of the color gauge group in the fundamental representation having the standard commutation relatios $[t_{a}, t_{b}] = i\,f_{abc}\,t_{c}$ and are normalized as $2\,tr t_{a}\,t_{b} = \delta_{ab}$; $ e$ is the strong coupling constant.
The leading tree level perturbative contribution is given by the form-factor $D_{2}$, while $D_{1}$ is different from zero and is dominated in IR region by decreasing behavior with the fall off controlled by a finite mass parameter $M$.
The parametrization of the form 
\begin{equation}
\label{e2}
D_{1}(x^{2}) = A\,e^{-M\,\vert\vec x\vert} + \frac{a}{x^{4}}\,e^{-\tilde M\,\vert\vec x\vert} + ... 
\end{equation}
obeys TPCF (\ref{e1}) at short and large distances. The parameters $A, a, M, \tilde M$ are estimated through the lattice (cooled) data (see, e.g., [9] and the refs. therein). 
At large distances the first term in  (\ref{e2})  becomes important. 

It has been shown [9] that $D_{1}$ is a nonvanishing function  if one assumes an effective dual approach to QCD where the main object is Abelian field strength $G_{\mu\nu} = \partial_{\mu} C_{\nu} - \partial_{\nu} C_{\mu}$, where $C_{\mu}$ is the dual vector potential. At large distances the dual to the dual field strength $\tilde G_{\mu\nu} =(1/2)\epsilon_{\mu\nu\alpha\beta}\,G_{\alpha\beta}$ behaves as the corresponding TPCF in QCD:
\begin{equation}
\label{e3}
\langle g^{2}\,\tilde G_{\mu\nu}(x)\,\tilde G_{\lambda\rho}(0)\rangle \sim
\langle e^{2}\,F_{\mu\nu}(x, x_{0})\,F_{\lambda\rho}(0,x_{0})\rangle, 
\end{equation}
where $g$ is the dual coupling constant, $x_{0}$ stands as an arbitrary reference point on the surface $S(\Gamma)$ (in the Wilson loop with the contour $\Gamma$) needed to surface ordering. 

{\it\bf Dilaton and scale symmetry breaking.-} In papers [10,11] there were introduced in some sense the observation of duality that means the following: many features of color confinement in QCD could be understood if continuum YM theory possessed some of properties of a magnetic superconductor. Thus, the dual potentials $C^{a}_{\mu}$, instead of $A^{a}_{\mu}$, are the natural variables to use in the confining regime.
In our model the duality means the assumption that physics of YM theory at large distances $r = \vert \vec x\vert$ depending upon strongly coupled gauge potential $A_{\mu}$ is the same as the  $r \rightarrow\infty$ physics of the dual theory describing the interactions of the potential $C_{\mu}$ weakly coupled to three scalar  fields $\phi_{i}\, (i=1,2,3)$ 
each in the adjoint representation of magnetic gauge group.
In four  dimensional gauge theories the field $\phi _{i}$ is associated with spontaneous breaking of (an approximate) scale symmetry. A light $CP$ -even scalar, a dilaton, may arize as a generic pseudo-Goldstone boson from the breaking of an approximate conformal strong dynamics [12]. 

We start with partition function
\begin{equation}
\label{e4}
Z = \int {\it\bold D}\phi_{i}\,\exp \left [-\int_{0}^{\infty}\,d\tau\,\int\, d^{3} x\,L(\tau, \vec x)\right ],
\end{equation}
where the integral is taken over fields periodic in Euclidean time  $\tau$ with period $\beta$  in thermal (heatbath) equilibrium at temperature $T = \beta^{-1}$. In general, the Lagrangian density (LD) in (\ref{e4}) given in the operator form
\begin{equation}
\label{e5}
L(x) = \sum_{i} c_{i}(\mu)\, O_{i} (x)
\end{equation}
is renormalized at the scale $\mu$; $c_{i}(\mu)$ is running coupling,  and the operator $O_{i} (x)$ has the scaling dimension $d_{i}$. Under the scale transformations $x^{\mu}\rightarrow e^{\omega}\,x^{\mu}$ one has $O_{i} (x)\rightarrow e^{\omega\,d_{i}}\,O_{i} (e^{\omega}x)$, $\mu\rightarrow e^{-\omega}\mu$. This gives for the divergence of the (scalar) dilatation current $S^{\mu} = T^{\mu\nu}x_{\nu}$ [12]
$$\partial_{\mu} S^{\mu} = T_{\mu}^{\mu} = \sum_{i} \left [c_{i}(\mu) (d_{i} - 4) O_{i}(x) + \beta_{i} (c) \frac{\partial}{\partial c_{i}} L\right ],$$
where $T^{\mu\nu}$ is the energy-momentum tensor, $\beta_{i} (c) = \mu\partial c_{i} (\mu) /\partial\mu$ is the running $\beta$-function. 

It is known that the gauge theories with non-trivial IR fixed point (IRFP) can describe the slow running of the strong coupling constant $\alpha$ that in some sense turns to the smallness of the $\beta$-function. For an $SU(N)$ theory with $N_{f}$ flavors in fundamental representation 
$$\beta (\alpha) = -\frac{b_{0}}{2\,\pi}\alpha^{2} -\frac{b_{1}}{(2\,\pi)^{2}}\alpha^{3} - ... ,$$
where $b_{0}$ and $b_{1}$ are known coefficients which can be found elsewhere.
The couplings $c_{i} (\mu)$ in (\ref{e5}) that in general are not scale invariant can be made scale invariant by introducing appropriate powers of the dilaton field to compensate for the shift under a scale transformations. At energies below the conformal breaking scale $f$ the coupling 
$c_{i} (\mu)$ has to be replaced [12]
$$c_{i}(\mu)\rightarrow \left (\frac{\phi}{f}\right )^{4-d_{i}}\,c_{i}\left (\mu\frac{\phi}{f}\right),$$
where the incorporated flat direction (the dilaton field) $\phi$ transforms according to $\phi (x)\rightarrow e^{\omega}\phi (e^{\omega}x)$ and $f = \langle \phi\rangle$ is the order parameter for scale symmetry breaking, determined by the dynamics of the underlying strong sector.
Actually, the theory would be nearly scale invariant if $d_{i} = 4$ and $\beta (\alpha)\rightarrow 0$. 
For an approximate dilatation symmetry (small $\beta (\alpha)$) or scale symmetry the IRFP $\alpha^{\star} = -2\pi\,b_{0}/b_{1}$  may have to be strong enough (with small $N_{f}$), however, it exceeds a critical strength $\alpha_{c}$ (with respective $N^{c}_{f}$) for the spontaneous breaking of chiral symmetry that leads to appearance of confinement. The scale of confinement associated with the scale of chiral symmetry breaking is of the order of the QCD scale $\Lambda\sim O(0.5\, GeV)$ at which $\alpha (\mu)$ crosses $\alpha_{c}$. 
Thus, the breaking of chiral symmetry is triggered by the dynamics of nearly conformal sector. 

The theory with fixed $\alpha^{\star}$ now possesses an exact scale invariance in the limit of chiral symmetry. The scale invariance is spontaneously broken when the chiral symmetry is also broken spontaneously [13]. As a result, a massless scalar particles, the dilatons - the Goldstone bosons of scale symmetry - appear. Since the actual gauge coupling constant is running and the scale symmetry is explicitely broken, the dilaton associated with the condensate should appear as a pseudo-Goldstone boson with a mass of the order of  $\Lambda$.



{\it\bf Critical mode.-}  Our aim is to study the critical  IR behavior of a scalar field theory in the presence of a random external sources. For  illustration of the critical mode we use the description of the fluctuations based on the partition function
$$ P[c] \sim \exp\{-\Delta H [c]\beta\}$$
which is the probability distribution of an order parameter field $\vec\varphi (x)$ coupled by an appropriate manner to the critical random field $\vec c(x)$ - the mode developing infinite fluctuation length at the (critical) ChEP. In the toy model for the phase transition in $D-$ dimensional coordinate isotropic space the effective action functional $\Delta H [c] \sim \int d^{D} x\,\vec c(x)\cdot\vec\varphi (x)$, and the scalar product is in the order-parameter space. We assume the random critical mode $\vec c (x)$ has a short-range spatial correlations $\hat\rho$
$$ \langle \vec c(x)\,\vec c(y)\rangle = \hat\rho (x-y), \,\,\,\langle \vec c_{k}\,\vec c_{k^{\prime}}\rangle = \delta_{k,k^{\prime}}\,\rho (k), $$
where $\vec c_{k} = V^{-1/2}\int d^{D} x \,\vec c(x) \,e^{ikx}$. The ChEP is characterized by $\hat\rho \rightarrow\infty$. In the dual gauge model considered in this paper, $\vec c(x)$ field is treated as the vector field, in particular, to bound two (color) charges in the condensate (the scalar dilaton field  $\vec\varphi (x)$).

In general, the distribution of the energy averaged over a random field $c(x)$ with the Gaussian shape is given by the integral
$$ F_{r} = \int {\it \bold D} c\,F [c]\,\exp \left [-\frac{1}{2} \int d^{D}x\,c^{2}(x)\right ],$$
where index $r$ means "random";
$$ F[c] =\ln\int {\it \bold D} \varphi\,\exp\left \{-\int d^{D} x \left [L_{0}(x) + c(x)\,\varphi (x)\right ]\right \}, $$
and LD $L_{0}$ will be defined below. 
Using the result of [14] the two-point Green's function is ($x = (\vec x,x_{4})$, $\tau$ has the sense of "stochastic time") 
\begin{equation}
\label{e51}
\lim_{\tau\rightarrow +\infty} {\langle\varphi (x,\tau)\varphi (0,\tau)\rangle}_{r} \sim \int {\it \bold D} c\,\varphi_{c} (x)\,\varphi_{c} (0)\,\exp\left [-\frac{1}{2}\int d^{D} y\,c^{2}(y)\right ], 
\end{equation}
where $\varphi_{c}$ is the solution of the differential stochastic Langevin equation
$$\frac{\partial\varphi (x,\tau)}{\partial\tau} = -\frac{\delta L_{0} [\varphi (x,\tau)]}{\delta\varphi (x,\tau)} + c(x,\tau)$$
in the limit $\tau\rightarrow +\infty$.
The relation  (\ref{e51}) bounds the Green's function of real (physical) fields in Euclidean space to stochastic fields $\varphi (x,\tau)$ depending on the random source $c(x,\tau)$.
We assume that $c(x,\tau)$ obeys the (auto)correlation condition ${\langle c(x,\tau)\,c(y,\tau^{\prime})\rangle}_{r} = 2\delta^{(D)} (x-y)\delta (\tau - \tau^{\prime})$ near the critical point. The results of [15] allow to conclude that the Green's functions of the stochastic equation 
$$ (\Delta - m_{\varphi}^{2} -\lambda_{\varphi}\varphi^{2})\varphi = c, \,\,\,\,\,\,\Delta \equiv \partial_{\mu}\partial^{\mu}$$ 
in $D$ dimensions near the critical point are the same as those generated by the LD 
$$L_{0}(x) = -\frac{1}{2}\varphi (x)\Delta\varphi (x) + \frac{1}{2} m_{\varphi}^{2}\varphi ^{2} + 
\frac{1}{4}\lambda_{\varphi} \varphi^{4}$$
in $D-2$ dimensions. This has been proved [14] in terms of equivalence of the $D$ dimensional spin system in a random external magnetic field with $D-2$ dimensional spin system in the absence of a magnetic field. The decreasing of the dimension in the system under consideratin when the stochastic random interaction is neglected is related with the hidden (super)symmetry of the associated stochastic equation.

{\it\bf Dual model.-} In $SU(3)$ gluodynamics the effective LD of the dual model  is 
\begin{equation}
\label{e6}
L_{eff} = -\frac{1}{4}G_{\mu\nu}G^{\mu\nu} + \sum_{i=1}^{3}\left [\frac{1}{2} 
{\vert D_{\mu}^{(i)}\phi_{i}\vert}^{2} - \frac{1}{4}\lambda_{\phi} \left (\phi_{i}^{2} - \phi_{{0}_{i}}^{2}\right)^{2}\right ],
\end{equation}
where 
$$G_{\mu\nu} = \partial_{\mu} C_{\nu} - \partial_{\nu}C_{\mu} - i\,g [C_{\mu},C_{\nu}],\,\,
 D^{(i)}_{\mu}\phi_{i} = \partial_{\mu}\phi_{i} - ig [C_{\mu},\phi _{i}],\,\, C_{\mu}(x) = \sum_{a=1}^{8} C^{a}_{\mu} (x)\,t_{a}.$$

The coupling of the dilaton to dual field $C_{\mu}$ which becomes heavy at the scale of chiral symmetry breaking is dictated by the appropriate group algebra relations.
In the $[U(1)]^{2}$ Higgs-like model, $D_{\mu}^{(i)} =\partial_{\mu} + i\,g\,\epsilon^{a}\,C_{\mu}^{a}$ is the covariant derivative acting on the scalar fields $\phi_{i}$, where $i=1,2,3$ and $a=3,8$. The $\epsilon$'s are the root vectors of the group $SU(3)$: $\vec\epsilon_{1} = (1,0)$,   $\vec\epsilon_{2} = (-1/2, -\sqrt {3}/2)$,  $\vec\epsilon_{3} = (-1/2, \sqrt {3}/2)$. The gauge fields $C_{\mu}^{a=3,8}$ are dual to the diagonal components of gluon fields   $A_{\mu}^{a=3,8}$. The scalar d.o.f. $\phi_{i}$  appear due to the compactness of the residual abelian gauge group $[U(1)]^{2}$ in the abelian projection $SU(3)\rightarrow [U(1)]^{2}$. The LD (\ref{e6}) is gauge invariant under changing the fields $C_{\mu}^{a}\rightarrow C_{\mu}^{a} + \partial_{\mu} \alpha^{a}$, where $\alpha^{a=3,8}$ are the parameters of the gauge transformation. The condition $\sum_{i=1}^{3} arg\,\phi_{i} = 0$ reflects the restriction of the phases of $\phi_{i}$.

The components of $G_{\mu\nu}$ define the magnetic field $\vec H$, $H^{k} = G^{0k}$, and 
the electric field $\vec\varepsilon $, $\varepsilon^{k} = (1/2)\epsilon_{klm} G^{lm}$. 
The dual Wilson loop [16]
$$U_{C}(x,y) =P\exp\left [i\,g\int_{y}^{x} dz^{\mu}\,C_{\mu} (z)\right ]$$
defines $C_{\mu} (x)$, where $U_{C}$ is invariant under gauge transformations 
$$C_{\mu} (x)\rightarrow\Omega^{-1}_{C}(x)\,C_{\mu}(x)\,\Omega_{C}(x) + \frac{i}{g}\,\Omega^{-1}_{C} (x)\,\partial_{\mu}\Omega_{C}(x)$$
and $\Omega_{C} (x)$ being an element of magnetic-color gauge group.
The scalar fields $\phi_{i}(x)$ with the order parameters 
$\langle \phi_{i} (x)\rangle =  \phi_{{0}_{i}}$, the vacuum expectation values (v.e.v.),  are associated with not individual particles but the subsidiary magnetically charged objects which cannot be observed experimentally. 
The color structure of $\phi_{{0}_{i}}$ is given by [17]
$$\phi_{{0}_{1}} = \frac {f}{\sqrt {2N}}\,J_{x}, \,\,\,\phi_{{0}_{2}} = \frac {f}{\sqrt {2N}}\,J_{y},\,\,\,\, \phi_{{0}_{3}} = \frac {f}{\sqrt {2N}}\,J_{z},$$
where $J_{j}$ are the three generators of the $N$ dimensional irreducible  representation of the three ($j$= $x$, $y$, $z$) dimensional rotation group corresponding to angular momentum $J = (N-1)/2$.
In the confinement phase the dual (magnetic-like) gauge symmetry is broken due to dual scalar mechanism, and all the particles become massive. The quanta of $C_{\mu}$ acquires a mass $m\sim gf$ via the dual Higgs-like mechanism, hence the dual theory is weakly coupled at distances $r > 1/mass$, where the denominator being either the mass of dual gauge quanta or the mass $m_{\phi}\sim\sqrt {2\lambda_{\phi}}\,f$
of the scalar field with the coupling constant $\lambda_{\phi}$. In the proximity of  ChEP,
$\lambda_{\phi}\rightarrow 0$, so the scalar field is treated as a classical one.  
Using the scheme with the partially conserved dilatation current  [18] where the dilaton mass is scaled by $\Lambda$  
$$ m_{\phi} \simeq \sqrt {1 - \frac{N_{f}}{N^{c}_{f}}}\,\Lambda, \,\,\,\, N_{f} \leq N^{c}_{f},$$
the v.e.v. $f$ can be  estimated through the decay constant $f_{\pi}$ of the $\pi$-meson 
$$f\simeq \frac{1}{0.3\,\sqrt {2\lambda_{\phi}}} \sqrt {1 - \frac{N_{f}}{N^{c}_{f}}}\, f_{\pi} $$
with $N_{f}\rightarrow N^{c}_{f}$ as $\lambda_{\phi}\rightarrow 0$.

The interaction between two charges is due to color magnetic current $J_{\mu} (x) \sim \partial^{\nu} G_{\mu\nu} (x)$ in the scalar (dilaton) condensate, where $G_{\mu\nu} = \partial_{\mu}\,C_{\nu} - \partial_{\nu}\,C_{\mu} +G^{s}_{\mu\nu}$. The $G^{s}_{\mu\nu}$ is the Dirac string tensor representing a moving line from charge $-g_{m}$ to the charge $+g_{m}$
$$G^{s}_{\mu\nu} (x) = g\,\epsilon_{\mu\nu\alpha\beta}\,\int_{0}^{1}d\tau\int_{0}^{1}d\sigma\,\frac{d y_{\alpha}}{d\sigma}\frac{d y_{\beta}}{d\tau}\,\delta^{4} [x - y (\tau,\sigma)],$$
where $y_{\mu} (\tau,\sigma)$ is a world sheet of a surface $S (\Gamma)$ swept by the Dirac string connecting $-g_{m}$ and $+g_{m}$. The fixed contour $\Gamma$ on the $S(\Gamma)$ depends on the charge source trajectories $z_{\mu}$: $\Gamma [z_{{1}_{\mu}} = y_{\mu}(\tau,\sigma =1),  z_{{2}_{\mu}} = y_{\mu}(\tau,\sigma =0)]$. The divergence of dual to Dirac string tensor is the current carried by a charge $g_{m}$ moving along the path $\Gamma$: $\partial^{\beta}\tilde G^{s}_{\alpha\beta} (x) \sim g\int dz_{\alpha}\,\delta ^{4} (x-z)$. The quantization condition $e\cdot g = 2\,\pi$ guarantees that the Dirac string will not be observable when the dual gauge symmetry is  not broken.

{\it\bf Flux tubes.-} The excitations above the (classical) vacuum in the effective theory are flux tubes connecting a quark-antiquark pair in which $Z_{N}$ electric flux is confined to narrow tubes of a radius $\sim m^{-1}$, at whose center the scalar condensate vanishes.
The probability distribution related to the ensemble of systems containing a single flux tube with $N(R)$ number of configurations of the flux tube of length $R$ is [19] 
\begin{equation}
\label{e7}
P = \sum_{\beta}\sum_{R}\,N(R)\,\exp \left [-\beta\,E (m,R)\right ]\,D (\vert \vec x\vert, \beta; M),
\end{equation}
where $E (m, R)$ is the effective action - the energy of the peace of the isolating string-like tube 
$$E (m, R)\sim m^{2}\,R [a + b\,\ln (\tilde\mu\,R)].$$
Here, $a$ and $b$ are known constants, $m$ is the mass of the critical mode $C_{\mu}$ which develops an infinite fluctuation  length $\xi\sim m^{-1}$ in the proximity of the ChEP with  the critical temperature $T_{c} = \beta_{c}^{-1}$. Note, that $m^{2}(\beta)\sim g^{2}(\beta)\,\delta^{(2)}(0)$, where $\delta^{(2)}(0)$ is the inverse cross-section of the flux tube, and could be expressed in terms of string radius $r_{s}$ as follows: $\delta^{(2)}(0)\sim c/(\pi\,r^{2}_{s})$, $c\sim O(1)$.
The function $D$ in (\ref{e7}) is associated with the scalar part of TPCF (\ref{e1}) in the form of large distances exponential fall-off of correlator of gauge-invariant operators $ O$ 
\begin{equation}
\label{e8}
\langle O(\tau,\vec x)\,O(\tau, 0)\rangle \sim A\,{\vert \vec x\vert}^{c}\, D(\vert \vec x\vert, \beta; M), \,\,\, as\,\,\, \vert\vec x\vert\rightarrow \infty, 
\end{equation}
$$D(\vert \vec x\vert, \beta; M) = \exp \left [- M(\beta)\,\vert\vec x\vert\right ], $$
where we also admit the existence of $D$ function at $T > T_{c}$; $c$ in (\ref{e8})  is the constant depending on the choice of the operator $O(\tau, \vec x)$. Within the dual conformity (\ref{e3}) we assume that $M^{-1}(\beta)$ is the measure of the screening effect of color electric field. In $SU(N)$ theory with $N = 2,3$ at hight $T$ and zero chemical potential with $N_{f}$ massless quark flavors $M(\beta)$ is [20]
$$M(\beta) = M^{LO}(\beta) + N\,\alpha\,T\,\ln\left [\frac{M^{LO}(\beta)}{4\,\pi\,\alpha\,T}\right ] + 4\,\pi\,\alpha\,T\,y_{n/p} (N) + O(\alpha^{2}\,T), $$
where 
$$M^{LO}(\beta) = \sqrt {4\,\pi\,\alpha\left (\frac{N}{3} + \frac{N_{f}}{6}\right )}\,T,$$
 $y_{n/p} (N) = c_{N} + d_{N,N} \sqrt{4\,\pi\,\alpha} $ and the coefficients $c_{N}$ and $d_{N,N}$ can be found in [20]. For large distances bounded by $\vert \vec x\vert < M^{-1} (\beta)$ at high temperatures the correlator  (\ref{e8}) is (for $c= -4$ with $V$ stands as the volume)
\begin{equation}
\label{e9}
\langle O(\tau,\vec x)\,O(\tau, 0)\rangle \sim -\frac{16\,A\,\pi}{3}\,\frac{T}{V}\,\sigma_{eff}(\beta)\,y_{n/p}(N)\,\xi^{2} 
\end{equation} 
that means the nonperturbative TPCF is explained in terms of fluctuation length $\xi$. The effective theory describes the fluctuations at distances larger that $\xi$ (or only at energy scales less than $m$) up to ChEP.
In the dual YM theory at finite temperature  the effective string tension in (\ref{e9})  is $\sigma_{eff}(\beta) \sim m^{2}(\beta)\,\alpha (\beta)$ [19]. 
In the scheme with the flux tubes, $\xi$ is the penetration depth of color-electric field (or, approximately,  the radius of the flux tube), while the dilaton mass inverse $l = m^{-1}_{\phi}$ stands for the coherent length of the scalar field (condensate). 
The scalar (dilaton) condensate ${\langle\phi\phi\rangle}_{\beta}$ at finite $\beta$ can be understood through the relation
\begin{equation}
\label{e10}
\frac{{\langle\phi\phi\rangle}_{\beta}}{2\,\lambda_{\phi}(\beta)} =\gamma_{conf}\,f^{2},
\end{equation} 
where 
$\gamma_{conf} = (2\,\pi)^{-3}\int d^{3}x\,(2\,\sqrt{x^{2} +1})^{-1}$ is the conformal factor. The l.h.s. of (\ref{e10}) is finite up to ChEP, where $\lambda_{\phi}(\beta = \beta_{c}) = 0$. The r.h.s.  of (\ref{e10}) tends to infinity if  $f^{2}\,x= \vec p^{2}/(2\,\lambda_{\phi})\rightarrow\infty$ as $\beta\rightarrow \beta_{c}$. 
The formation time of the flux tube is $\tau = \sqrt {4/(3\alpha)}\xi$ which tends to infinity at ChEP.  For $SU(3)$, $m\simeq 2\sqrt {\sigma_{eff}}$ [21], and the lattice simulations yield $T_{c}\simeq 0.65 \sqrt {\sigma_{eff}}$. Thus, the effective theory should also be applicable in the deconfined phase in the range of temperatures within the interval $T_{c} < T < 3 T_{c}$. 
 

The interactions between flux tubes are defined by the scalar and gauge boson field profiles. 
 The ratio $k = \xi/l$ as the Ginzburg-Landau-like parameter defines the properties of the dual superconductor QCD vacuum. The attracted forces can appear between two (parallel) flux tubes (of the same type) if $k < 1$ (type-I vacuum), otherwise the flux tubes repel each other in the vacuum where $k >1$ (type-II vacuum), and ChEP is characterized by $k\rightarrow \infty$. In the case $k < 1$ the attraction is due to exchange of the scalar boson  which prevails over a gauge boson attraction. The interaction between (parallel) flux tubes of different types is complicated, however, these interactions were discussed in [22] (see also [23]) due to important aspects owing to Weyl symmetry.

{\it\bf Two-particle correlations.-} 
The physical context is important for the procedure of quantization of causal random (stochastic) processes. In physics, the random processes serve for the description of the behavior of open systems which are restricted and interact with (infinite) environment. The latter serve as to support the constant temperature. An open system accompanied by the thermostat is the complete system where any dynamical fluctuations might be occurred.
 
The fluctuations of some modes in the vicinity of ChEP can not be measured in experiments. These fluctuations can affect the fluctuations of observable either in direct channel or in the indirect reactions. One of the examples is the two-particle BE correlation function where the strength of the correlation between two particles may have the influence fluctuations.

The two-particle BE correlation function 
\begin{equation}
\label{e11}
\left\langle C_{2} (q, P) = \frac{\int dx_{1}\,dx_{2}\,S (x_{1},p_{1})\,S (x_{2},p_{2}){\vert \Phi\vert}^{2}}
{\int dx_{1}\,S (x_{1},p_{1})\,\int dx_{2}\,S (x_{2},p_{2})}\right \rangle_{\beta}
\end{equation} 
in the sense of complete system is related to the emission function $S(x,P)$ (of two particles with four-momenta $p_{1}$ and $p_{2}$) which is the Wigner-like thermalized phase-space density of the particle emitting system and can be viewed as the probability that a particle with average momentum $P = (p_{1} + p_{2})/2$ is emitted from the space-time point $x$ in the collision region; $q = p_{1} - p_{2}$; $\Phi$ is in general the two-particle state function. The index $\beta$ reflects the dependence of $C_{2}$ on a temperature. 

At finite temperatures, the function $C_{2}$  (\ref{e11})  has the form  [7]
\begin{equation}
\label{e12}
C_{2} (q,\beta) \simeq \eta (n)\left\{1+ \tilde\lambda(\beta) e^{-q^{2}L^{2}_{st}} \left [ 1 + \lambda_{1}(\beta) e^{+q^{2}L^{2}_{st}/2}\right ]\right \},
\end{equation} 
where for $n$ particle multiplicity $\eta (n) = \langle n(n-1)\rangle/\langle n\rangle^{2}$; $L_{st} = L_{st}(\beta, m)$ is the measure of the space-time overlap between two identical particles affected by stochastic forces. The special attention has to be paid to $\tilde\lambda$ and $\lambda_{1}$ in formula (\ref{e12}). In the standard quantum-mechanical scheme, e.g. Goldhaber-like one [24],  $\tilde\lambda$ is the positive $c$-number restricted by  1, while $\lambda_{1} = 0$. 
Near the ChEP the strength of BE correlation between two particles $\tilde\lambda (\beta)$ vanishes with a power of $\xi$ given by
\begin{equation}
\label{e13}
\tilde\lambda (\beta) = \lambda (\beta)\vert_{\xi = L_{st}}\,L_{st}^{2}\,\frac{1}{\xi^{2}},
\end{equation} 
where [19]
$$\lambda (\beta) =\frac{\gamma (\omega,\beta)}{ (1 + \nu)^{2}},\,\, \lambda_{1} (\beta) = \frac{2\,\nu}{\sqrt {\gamma(\omega,\beta)}}$$
and the stochastic forces influence is given by  $\nu = \nu (\omega,\beta) < \infty $. 
The function $\gamma (\omega,\beta)$ calls for quantum thermal properties of the particle source:
$$ \gamma (\omega,\beta) = \frac{\hat n^{2} (\bar\omega)}{\hat n(\omega)\,\hat n(\omega^{\prime})}, \,\,\, \hat n(\omega) = \frac{1}{e^{(\omega - \mu)\beta} -1}, \,\,\, \bar\omega = \frac{\omega + \omega^{\prime}}{2}, $$
$\omega$ is the energy of the particle with momentum $p = (\omega,\vec p)$ in thermal bath with statistical equilibrium at the temperature inverse $\beta$; $\mu$ is the chemical potential. If the vacuum of $SU(3)$ gluodynamics lies near a border between type-I and type-II superconductivity ($k\simeq 1$) the fluctuations of $\tilde\lambda$ in (\ref {e13}) are defined by vacuum condensate with the fluctuation length $\sim\xi$. In the border where $k = 1$ the parallel strings which carry the same flux do not interact with each other. Thus, the study and observation of the correlations between two bound states (in terms of strings) allocated in thermal bath are rather useful to check one is approaching to ChEP. 

{\it \bf Results on qualitative level.-}

1.  {\it Dip-effect.}
At low temperatures, $T < \sqrt {m^{2}_{h} + \mu^{2}}$, in the case of light hadrons with the mass $m_{h}$ the function $C_{2}$ can approach even below unity (dip-effect) in the small region of $q$ with $\langle n\rangle \sim O(10)$, 
where the (correlated) system defined by evolving stochastic scale $L_{st}$
\begin{equation}
\label{e14}
L_{st}\simeq \left [\frac{(2\,\pi)^{3/2}\,e^{\sqrt {m^{2}_{h} + \mu^{2}}\,\beta}}{3\,\nu (n)\,k^{2}_{T}\,(m^{2}_{h} + \mu^{2})^{3/4}\,T^{3/2}\,\left (1 + \frac{15}{8}\frac{T}{\sqrt {m^{2}_{h} + \mu^{2}}}\right )}\right ]^{1/5}
\end{equation} 
is disturbed by external force strength $\nu (n)$. 
In formula (\ref{e14}), the pair average transverse momentum $k_{T}$ is defined as half of the absolute vector sum of the two transverse momenta, $k_{T} = \vert \vec p_{T_{1}} + \vec p_{T_{2}}\vert /2$; $\vert \vec p_{T_{i}}\vert = \sqrt {\vec p^{2}_{x_{i}} + \vec p^{2}_{x_{i}}}$. The dependence of BE correlation signal on $k_{T}$ has been observed at the  SPS [25], at the Tevatron [26] and at RHIC [27]. The condition $\mu < m_{h}$ is a general restriction in the relativistic "Bose-like gas", while $\mu = m_{h}$ corresponds to  BE condensation. It is clearly seeing from (\ref{e14}) that the ChEP is a well-defined singularity on $L_{st}$.
 
The dip-effect,  unlike the monotonic shape of $C_{2}$ with asymptotic behavior as $C_{2} (q\rightarrow\infty) =1$, has already been observed (as the anticorrelation effect) in CMS experiment at the LHC [28] at $q\simeq [0.5 - 1.5]$ GeV depending on charged multiplicity $n$. The same qualitative behavior has been found earlier in L3 experiment at LEP [29]. At high temperatures  
the dip-effect disappears as $\langle n\rangle >> 1$, and at the ChEP the $C_{2}$ function does not deviate from 1. 

2. {\it Evolving size.} The size of the particle emission source in terms of stochastic scale $L_{st}$ is strongly dependent on $\nu (n)$, $k_{T}$, $m_{h}$ and $T$. We have shown in (\ref{e14}) that at low temperatures 
$L_{st}$ decreases with $k_{T}$ and increases with $n$ ($\nu (n)\rightarrow 0$), where 
\begin{equation}
\label{e15}
\nu (n) = \frac{2 - \tilde C_{2}(0) + \sqrt {2 - \tilde C_{2}(0) }}{\tilde C_{2}(0) -1},\,\, \,\,\,\tilde C_{2}(0) =\frac{C_{2}(q=0)}{\eta (n)} 
\end{equation}
with $\langle n\rangle \geq 1 + C_{2}(0) /2$. From theoretical point of view $C_{2} (0)$ can not exceed the value of 2.

At higher temperatures there is a nontrivial singular behavior of $L_{st}$ with $\nu (n)$ 
$$L_{st} \sim \left [\nu (n)\,k_{T}^{2}\,T^{3}\right ]^{-1/5}, $$
where no dependence of the chemical potential and the particle mass are found, and $L_{st}\rightarrow\infty$ as $\nu (n)\rightarrow 0$ with $n\rightarrow\infty $ (see (\ref{e15})). The CMS experiment [28] has confirmed that the effective emission radius increases with a charged-particle multiplicity in the events with colliding energies from  $\sqrt {s}$ = 0.9 TeV to 7 TeV, and it decreases smoothly with $k_{T}$. The effect of decreasing will be more sufficient in events with large particle multiplicities.
At the temperatures close to ChEP we can expect the strengthening of the expansion of the particle emission size with particle  multipilicity.  

3. {\it Correlation strength.} We find (see (\ref{e13}) and (\ref{e14})) that the correlation strength $\tilde\lambda (\beta)$ decreases with $k_{T}$. One can suppose that $\tilde\lambda$ is nearly independent of  $n$ at the position corresponding to $k_{T}$ where the main dependence on $n$ is given by $L_{st}$ in the ratio $L_{st}^{2}/[1 + \nu (n)]^{2}$. Actually, $\tilde\lambda (\beta)\rightarrow 0$ as $T\rightarrow T_{c}$ with $\xi\rightarrow\infty$. The result of smooth decreasing of $\tilde\lambda$ with $k_{T}$ with slight increasing of the effect (the value of $\lambda$) at low $n$ was  demostrated by CMS (see Fig.3 in [28]).


{\it\bf Conclusions.-}  To conclude, we were faced the ChEP phenomenon starting from first principles of strong interacting theory. The approach to this is based on $SU(N)$ YM theory described by an effective model coupling magnetic $SU (N)$ gauge potentials $C_{\mu}$ to 3 adjoint scalar fields, the dilatons. These couplings generate color magnetic currents which, via a dual Meissner effect, confine $Z_{N}$ electric flux to narrow tubes connecting a quark-antiquark pair. The dilaton associated with the condensate may be difficult  to detect as it has vacuum quantum numbers.

We studied the effect of ChEP on correlation strength $\tilde\lambda$ of fluctuations of BE correlation function $C_{2}$ as the experimental observable in heavy-ion collisions. The characteristic signature is non-monotonous  and even singular behavior of $C_{2}$ as a function of $\tilde\lambda$, which is sensitive to the proximity of ChEP and is measured by the magnitude of fluctuation length $\xi$.  The $C_{2}$ being the function of  $\sqrt {s}$, $T$, $k_{T}$, $\nu(n)$ has a trivial behavior defined by $\eta (n)\sim O(1)$ as the ChEP is approached and then passed. 


Finally, in this paper we have proposed:\\
a) the dip-effect may occur in the behavior of $C_{2}$ at small $\langle n\rangle$ at low temperatures;\\
b) at higher $T$ the dip-effect disappears as $\langle n\rangle >> 1$ and $C_{2}$ function does not deviate from 1;\\
c) the size $L_{st}$ of the particle emission source increases with $n$ at lower $T$;\\
d) at $T\rightarrow T_{c}$ the strengthening of the size $L_{st}$ is due to $\nu (n)\rightarrow 0$ and $m_{h}\rightarrow 0$;  the $L_{st}$ has a singularity at the transition point, ChEP; a phase transition appears only in the limit $L_{st}\rightarrow\infty$, but not with finite $L_{st}$;\\
e) the correlation strength $\tilde\lambda$ decreases with $k_{T}$; $\tilde\lambda$ is nearly independent of $n$ and $\tilde\lambda \rightarrow 0$ as $T\rightarrow T_{c}$ with $\xi\rightarrow\infty$.

The results a), c), e) as well as the disappearing of the dip (anticorrelation)-effect at $\langle n\rangle >> 1$ were already confirmed by CMS experiment [28]. The results b), d) and e) are the subjects of search either at NICA or at FAIR.

\end{document}